\documentclass[10pt,journal]{IEEEtran}
\usepackage{amsmath}
\usepackage{epsfig}
\usepackage{graphicx}
\usepackage{amsmath}
\usepackage{amsfonts}
\usepackage{latexsym}
\usepackage{amssymb}
\usepackage{cite}
\usepackage{array}

\newcommand{\br}{\mathbf{r}}
\newcommand{\bs}{\mathbf{s}}
\newcommand{\bn}{\mathbf{n}}

\newcommand{\bx}{\mathbf{x}}
\newcommand{\bc}{\mathbf{c}}
\newcommand{\by}{\mathbf{y}}

\newcommand{\bO}{\mathbf{0}}

\newcommand{\bA}{\mathbf{A}}

\newcommand{\bB}{\mathbf{B}}

\newcommand{\bH}{\mathbf{H}}

\newcommand{\bR}{\mathbf{R}}
\newcommand{\bW}{\mathbf{W}}

\newcommand{\bQ}{\mathbf{Q}}

\newcommand{\bC}{\mathbf{C}}

\newcommand{\bI}{\mathbf{I}}
\newcommand{\bP}{\mathbf{P}}

\newcommand{\bh}{\mathbf{h}}

\begin{document}

\title{Efficient Soft-Input Soft-Output MIMO Chase Detectors for arbitrary number of streams}
\author{Ahmad Gomaa and Louay M.A. Jalloul,~\IEEEmembership{Senior~Member,~IEEE}}

\maketitle

\begin{abstract}
We present novel soft-input soft-output (SISO) multiple-input multiple-output (MIMO) detectors based on the \emph{Chase detection} principle \cite{Chase_detectorsTSP08} in the context of iterative and decoding (IDD). The proposed detector complexity is linear in the signal modulation constellation size and the number of spatial streams. Two variants of the SISO detector are developed, referred to as SISO B-Chase and SISO L-Chase. An efficient method is presented that uses the decoder output to modulate the signal constellation decision boundaries inside the detector leading to the SISO detector architecture. The performance of these detectors significantly improves with just a few number of IDD iterations. The effect of transmit and receive antenna correlation is simulated. For the high-correlation case, the superiority of SISO B-Chase over the SISO L-Chase is demonstrated.

\end{abstract}

\setlength{\textfloatsep}{10pt}
\setlength{\floatsep}{8pt}

\section{Introduction}\label{sec_Introduction}
Multiple-input multiple-output (MIMO) antenna communication systems are known to achieve large spatial multiplexing and diversity gains in multi-path rich fading channels. All communication systems require some sort of error correction coding for reliable reception, e.g. turbo codes or low-density parity-check codes. The \emph{turbo principle} as described in \cite{Douillard95} has been successfully applied and extended to coded MIMO systems with iterative detection and decoding (IDD) \cite{Tuchler02,MAP-approx-Hassibi}. The "outer" code is the turbo or LDPC code and the "inner" code is spatial multiplexing and transmission of the symbols over the multiple antennas. With IDD there is an iterative exchange of information between the MIMO detector and the channel decoder which has been shown to achieve near channel capacity \cite{Hochwald03}. In IDD architectures, the log-likelihood ratios (LLRs) of the code bits are generated by the MIMO detector and passed to the channel decoder, which computes the extrinsic LLRs and feeds them back to the detector. The detector exploits the a priori LLRs from the decoder to generate more accurate LLRs for the channel decoder to start a new iteration.

There is a rich body of literature on MIMO (or symbol) detectors ranging in complexity from linear minimum mean-squared error (MMSE) \cite{Tuchler02, 2003_Wubben} to sphere detection \cite{2002_Agrell}. These detectors also differ in terms of their relative performance where a sphere detector typically performs close to maximum likelihood (ML) whereas the linear MMSE is inferior to ML. The core receiver complexity remains to be dominated by the MIMO detector especially when the number of spatial streams and the the size of the signal modulation constellation on each stream are large, e.g. 4-layers and 64-QAM.

The drawback of MIMO sphere detectors has been their hardware implementation complexity \cite{2004_Garret}. In spite of recent advances in complexity reduction of MIMO sphere detectors, as reported in \cite{2004_Wang_WCNC,2005_Yee_ICASSP,2008_Studer}, their nondeterministic processing throughput remains to present a challenge for hardware implementation \cite{2012_studer_scheduler}.

In this paper, we develop efficient soft-input soft-output MIMO detectors in the context of IDD. Our proposed MIMO detectors have linear complexity. Furthermore, their structures are generalizable to an arbitrary number of spatial streams and signal modulation constellation size and, moreover, have fixed detection times. Our proposed MIMO detectors are generalization of the Chase family of detection algorithms \cite{Chase_detectorsTSP08} (referred to in the sequel as SISO B-Chase and SISO L-Chase). Our contribution over the approach in \cite{Chase_detectorsTSP08} is two-fold: First, in our proposed SISO B-Chase detector, we account for the residual error variance in the symbol detection, making the detector capable of generating soft output information, unlike the B-Chase detector in \cite{Chase_detectorsTSP08} where only hard output is available. Second, we extend both of the L-Chase and B-Chase detection algorithm to use soft-input through an efficient method for processing the a priori information obtained from the decoder. We compare our detection method to that of \cite{SIOF2010} that uses soft feedback detection. We show large performance gains with the new detection method over iterative SIOF.

The rest of the paper is organized as follows. Section \ref{sec_SysModel} introduces the system model. Our soft-input soft-output MIMO detection methods are described in Sections \ref{sec_L_Chase} and \ref{sec_B_Chase} and their performances are shown in Section \ref{sec_simulations}. Finally, Section \ref{sec_Conclusion} has the concluding remarks.

\textit{Notations}: Unless otherwise stated, lower case and upper case bold letters denote vectors and matrices, respectively, and $\bO$ denotes the all-zero column vector. $\bI_m$ denotes the identity matrix of size $m$. The $k$-th element of the vector $\bx$ is denoted by $\bx(k)$. Furthermore, $\left|\;\right|$, $\|\;\|$, and $E[.]$ denote the absolute value, $l_2$-norm, and statistical expectation, respectively, while $(\,)^H$ denotes the complex conjugate transpose operation.

\section{System Model}\label{sec_SysModel}
We consider MIMO systems, where $N_L$ QAM symbols are linearly precoded using the precoding matrix $\bW$ of size $N_t\times N_L$ and then transmitted over $N_t\geq N_L$ antennas. The receiver detects the transmitted symbols (streams) using $N_r \geq N_L$ receive antennas. The input-output relation is given by
\begin{equation}\label{eqn_I/P_O/P}
\by = \bar{\bH} \bW \bs +\bn \triangleq \bH\bs+\bn= \sum_{i=1}^{N_L} \bh_i s_i +\bn
\end{equation}
where $\by$, $\bs$, $\bn$ and $\bar{\bH}$ denote the $N_r\times 1$ received signal, $N_L\times 1$ transmitted symbols, $N_r \times 1$ background noise plus inter-cell interference, and $N_r \times N_t$ channel matrix, respectively. Furthermore, $\bh_i$ is the $i$-th column vector of the equivalent channel matrix $\bH=\bar{\bH}\bW$, and $s_i$ is the $i$-th transmitted $M$-QAM symbol chosen from constellation $\chi$. The symbol, $s_i$, represents $q=\text{log}_2(M)$ code bits $\bc_i=\begin{bmatrix}c_{i1}&c_{i2}&\ldots& c_{iq}\end{bmatrix}$. The above relationship models both single-carrier systems over flat fading channels and orthogonal frequency division multiplexing (OFDM) systems over frequency-selective channels where Equation \eqref{eqn_I/P_O/P} applies to each subcarrier. Assuming a known channel at the receiver and zero-mean circularly symmetric complex Gaussian noise $\bn$ of covariance matrix $\bC_{\bn\bn}=\bB^{-1}$, we write the max-log maximum-a-posteriori (MAP) detector LLR of the bit $c_{ik}$ as follows \cite{Hochwald03,MAP-approx-Hassibi}:
\begin{align}
L(c_{ik})&=\max\limits_{s_i\in \chi_{k,1}} \eta_\text{max-log}(\bs) - \max\limits_{s_i\in \chi_{k,0}} \eta_\text{max-log}(\bs)\label{eqn_LLR_MaxLog}\\
\eta_\text{max-log}(\bs) &= \sum_{m=1}^{N_L} \sum_{n=1}^q b_{mn}L_a(c_{mn}) \hspace{-0.1cm}- \|\by\hspace{-0.1cm}-\hspace{-0.1cm}\sum_{m=1}^{N_L} \bh_m s_m \|_\bB^2
\end{align}
where $\bs=[s_1\,s_2\,...\,s_{N_L}]^T$, $\|\bx \|_\bB^2\equiv\bx^H\bB\bx$, and $\chi_{k,1}$ and $\chi_{k,1}$ denote the constellation sets where the $k$-th bit is '1' and '0', respectively. Furthermore, $L_a(c_{mn})$ denotes the a priori LLR (computed by the decoder) of the bit $c_{mn}$, while $\left\{b_{mn}\right\}_{n=1}^q\in\left\{0,1\right\}$ denote the bit vector representation of $s_m$. The exact brute-force solution of \eqref{eqn_LLR_MaxLog} requires the computation of $M^{N_L}$ metrics, which is quite complex for large signal modulation constellation sizes and large number of spatial streams. However, in this work we show how we approximate the max-log MAP solution in \eqref{eqn_LLR_MaxLog} using much lower complexity. By whitening the noise and multiply $\by$ by $\sqrt{\bB}$, we arrive at an equivalent system model given by
\begin{align}
\tilde{\by}=\sqrt{\bB}\by=\widetilde{\bH}\bs+\tilde{\bn}
\end{align}
where $\widetilde{\bH} = \sqrt{\bB} \bH$ and $ \tilde{\bn} = \sqrt{\bB} \bn \sim N(\bO,\bI_{N_r})$.

\section{Soft-Input Soft-Output L-Chase Detectors}\label{sec_L_Chase}
In order to generate the LLRs of $s_i$, we first reorganize the columns of $\widetilde{\bH}$ such that its $i$-th and last columns are swapped to get $\bH_i=\widetilde{\bH}\bP_i$, where $\bP_i$ is the corresponding $N_L \times N_L$ permutation matrix with $\bP_i^2=\bI_{N_L}$. Next, we obtain the QR decomposition of $\bH_i\equiv\bQ_i \bR_i$ and rotate $\by$ by $\bQ_i^H$ to get
\begin{align}
\by_i = \bQ_i^H \tilde{\by} = \bR_i \bP_i \bs + \bn_i
\end{align}
where $\bQ_i^H\bQ_i=\bI_{N_L}$, $\bn_i=\bQ_i^H\tilde{\bn}\sim N(\bO,\bI_{N_L})$ and $\bR_i$ is an $N_L\times N_L$ upper triangular matrix. Furthermore, $\bP_i \bs=[s_{j_1}\,s_{j_2}\,...\,s_{j_{N_L}}]^T$ where $j_{N_L} = i$ and $j_i=N_L$. The matrix $\bR_i$ can be portioned as follows:
\begin{align}
\bR_i\equiv\begin{bmatrix} \widetilde{\bR}_i & \tilde{\br}_i  \\ \bO^T & d_i\end{bmatrix}
\end{align}
where $\widetilde{\bR}_i$ is the upper triangular $(N_L-1)\times(N_L-1)$ matrix constructed by the first $N_L-1$ rows and columns of $\bR_i$. Also, $\tilde{\br}_i$ is the last column of $\bR_i$ excluding its $(N_L,N_L)$ element denoted by $d_i$. Next, we get
\begin{align}\label{eqn_I/P_O/P_linear}
\bar{\by}_i \hspace{-0.1cm}= \begin{bmatrix} \widetilde{\bR}_i^{-1} & 0  \\ 0 & 1\end{bmatrix} \by_i =
\begin{bmatrix} \bI_{N_L-1} & \bc_i  \\ \bO^T & d_i\end{bmatrix} \bP_i \bs + \bar{\bn}_i
\end{align}
where $\bc_i = \widetilde{\bR}_i^{-1} \tilde{\br}_i$ is obtained using the back substitution technique \cite{Golubbook} where no matrix inversion is needed. Using the new input-output relation in \eqref{eqn_I/P_O/P_linear}, we approximate the LLR in \eqref{eqn_LLR_MaxLog} as follows:
\begin{align}
L(c_{ik})&\approx\max\limits_{s_i\in \chi_{k,1}} \eta_\text{L-Chase}(\bs) - \max\limits_{s_i\in \chi_{k,0}} \eta_\text{L-Chase}(\bs)\label{eqn_LLRmax_LChase}\\
\eta_\text{L-Chase}(\bs) \hspace{-0.1cm}&=\hspace{-0.1cm} \sum_{m=1}^{N_L} \sum_{n=1}^q b_{mn}L_a(c_{mn}) \hspace{-0.1cm}- \hspace{-0.1cm}\left\|\bar{\by}_i-\hspace{-0.1cm}\begin{bmatrix} \bI_{N_L-1} & \bc_i  \\ \bO^T & d_i\end{bmatrix} \bP_i \bs\right\|^2 \nonumber\\
&=\hspace{-0.1cm} \sum_{m=1}^{N_L} \sum_{n=1}^q b_{mn}L_a(c_{mn}) \hspace{-0.1cm}- \left|\bar{\by}_i(N_L)-d_is_i\right|^2 \nonumber\\ &-\sum_{l=1}^{N_L-1} \frac{\left|\bar{\by}_i(l)-\bc_i(l)s_i-s_{j_l}\right|^2}{E[\left|\bar{\bn}_i(l)\right|^2]} \label{eqn_LLR_LChase}
\end{align}
where $E[\left|\bar{\bn}_i(1)\right|^2]=1$ and $E[\left|\bar{\bn}_i(l)\right|^2]$ is the $l_2$-norm of the $l$-th row of $\widetilde{\bR}_i^{-1}$. Note that the correlations between the elements of $\bar{\bn}_i$ are not accounted in \eqref{eqn_LLR_LChase} in order to reduce the algorithm complexity. We rewrite the maximization problems in \eqref{eqn_LLRmax_LChase} as:
\begin{align}
&\max\limits_{s_i\in \chi_{k,\hspace{-0.05cm}1(0)}} \hspace{-0.15cm} \eta_\text{L-Chase}(\bs) \hspace{-0.1cm} = \hspace{-0.3cm} \max\limits_{s_i\in \chi_{k,\hspace{-0.05cm}1(0)}} \hspace{-0.2cm} \Bigg(\hspace{-0.05cm}\underbrace{\sum_{n=1}^q b_{in}L_a(c_{in}) \hspace{-0.1cm}-\hspace{-0.1cm} \left|\bar{\by}_i(N_L)\hspace{-0.1cm}-\hspace{-0.1cm}d_is_i\right|^2}_{\triangleq \alpha_i}  \nonumber\\
&+\hspace{-0.1cm}\sum_{l=1}^{N_L-1}\hspace{-0.1cm} \underbrace{\max\limits_{s_{j_l}\in \chi} \hspace{-0.1cm} \left(\sum_{n=1}^q b_{j_l n}L_a(c_{j_l n}) \hspace{-0.1cm} - \hspace{-0.1cm} \frac{\left|\bar{\by}_i(l)-\bc_i(l)s_i\hspace{-0.1cm}-\hspace{-0.1cm}s_{j_l}\right|^2}{E[\left|\bar{\bn}_i(l)\right|^2]}\right)}_{\triangleq \alpha_{i,l}}\hspace{0.05cm}\Bigg)
\label{eqn_max_problem_LChase}\end{align}
where we enumerate over the stream of interest $s_i$, and compute $\alpha_i+\sum_{l=1}^{N_L-1}\alpha_{i,l}$ for every possible instance of $s_i$. Then, we run the maximization over $\chi_{k,1}$ and $\chi_{k,0}$ to get the LLR. Although \eqref{eqn_max_problem_LChase} reveals that we need to compute $(N_L-1)M^2$ metrics, we show in the next subsection how we \textit{exactly} and efficiently solve for the sub-maximization problem $\alpha_{i,l}$ using much fewer metric computations.

\subsection{Efficient computation of the sub-maximization problem}\label{sec_maximiz_comput}
\begin{figure}
\centering
\epsfig{file=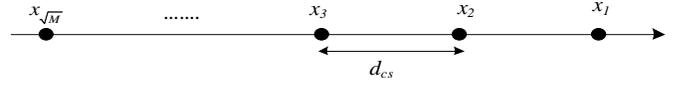,height=0.5in,width=3.45in}
\caption{$\sqrt{M}$-PAM constellation $\psi$ for the real (or imaginary) part of the $M$-QAM constellation $\chi$.} \label{fig_constellation}
\end{figure}

Separating out the real and imaginary parts of the sub-maximization problem $\alpha_{i,l}$, we write \cite{ExactMaxLog_CL14}
\begin{align}\label{eqn_submax_prob}
\alpha_{i,l}&=\max\limits_{s_{r,j_l}\in \psi}  \left(\sum_{n\in I_r} b_{j_l n}L_a(c_{j_l n})  -  \frac{\left(z_{r,l}-s_{r,j_l}\right)^2}{E[\left|\bar{\bn}_i(l)\right|^2]}\right) \nonumber\\
&+\max\limits_{s_{I,j_l}\in \psi}  \left(\sum_{n\in I_I} b_{j_l n}L_a(c_{j_l n})  -  \frac{\left(z_{I,l}-s_{I,j_l}\right)^2}{E[\left|\bar{\bn}_i(l)\right|^2]}\right)
\end{align}
where $z_{r,l}(s_{r,j_l})$ and $z_{I,l}(s_{I,j_l})$ denote the real and imaginary parts of $z_l(s_{j_l})$, respectively, where $z_l=\bar{\by}_i(l)-\bc_i(l)s_i$. Furthermore, $I_r$ and $I_I$ are the bit indices corresponding to $s_{r,j_l}$ and $s_{I,j_l}$, respectively, while $\psi$ denotes the $\sqrt{M}$-PAM one-dimensional constellation shown in Fig. \ref{fig_constellation} which corresponds to the real or imaginary parts of $s_{j_l}$.
We use the a priori LLRs to modulate \cite{ExactMaxLog_CL14} the decision thresholds between the $\sqrt{M}$-PAM constellation symbols in Fig. \ref{fig_constellation}. Based on the modulated thresholds, we apply simple slicers on $z_{r,l}$ and $z_{I,l}$ to obtain the solutions of the real and imaginary maximization problems in \eqref{eqn_submax_prob} denoted by $s_{r,j_l}^*$ and $s_{I,j_l}^*$, respectively, as:
\begin{align}
s_{r,j_l}^* \hspace{-0.1cm}&=\hspace{-0.05cm} \text{arg}\max\limits_{s_{r,j_l}\in \psi} \hspace{-0.1cm}  \left(\sum_{n\in I_r} b_{j_l n}L_a(c_{j_l n})  -  \frac{\left(z_{r,l}-s_{r,j_l}\right)^2}{E[\left|\bar{\bn}_i(l)\right|^2]}\right)\\
&=Q[z_{r,l}] \triangleq
\begin{cases} x_1, &  \max\limits_{u>1} D_{1u} \leq z_{r,l} \\
 x_m, &  \max\limits_{u>m} D_{mu} \leq z_{r,l} \leq  \min\limits_{u<m} D_{mu} \\
 x_{\sqrt{M}}, &   z_{r,l} \leq  \min\limits_{u<\sqrt{M}} D_{\sqrt{M}u}
\end{cases}\nonumber
\end{align}
where $Q[.]$ is the quantization (slicing) function and\footnote{For the detailed derivation, refer to \cite{ExactMaxLog_CL14}.}
\begin{align}
D_{mu}&=D_{um}=\frac{x_m+x_u}{2}\nonumber\\
&-E[\left|\bar{\bn}_i(l)\right|^2]\frac{\sum_{n\in I_r} (b_{m n}-b_{u n})L_a(c_{j_l n})}{2(x_m-x_u)}
\end{align}
is the modulated boundary between the one-dimensional constellation symbols $x_m$ and $x_u$ in Fig. \ref{fig_constellation}, where $\left\{b_{mn}\right\}_{n=1}^q$ is the bit vector representation of any $M$-QAM complex symbol whole real part equals to $x_m$. Similarly, we obtain $s_{I,j_l}^*$. Despite that the solution of \eqref{eqn_max_problem_LChase} requires enumeration over $s_i$, the $M-\sqrt{M}$ boundaries, $\left\{D_{mu}\right\}$, are computed only once and not for each instance of $s_i$. Hence, the total number of metric computations required to detect $s_i$ is $N_LM\hspace{-0.05cm}-\hspace{-0.05cm}(N_L-1)\sqrt{M} \ll (N_L-1)M^2$ for large $N_L$ and $M$.

\section{Soft-Input Soft-Output B-Chase Detectors}\label{sec_B_Chase}
To detect the $i$-th stream, we reorganize the columns of $\widetilde{\bH}$ such that its $i$-th column is moved to the last column position. The remaining $N_L-1$ columns are BLAST-sorted \cite{VBLAST} and placed at the first $N_L-1$ column positions. The reorganized channel matrix is $\bH_i=\widetilde{\bH}\bA_i$, where $\bA_i$ is the $N_L \times N_L$ permutation matrix with $\bA_i^2=\bI_{N_L}$. Next, we obtain the QR decomposition of $\bH_i\equiv\bQ_i \bR_i$ and rotate $\by$ by $\bQ_i^H$ to get
\begin{align}
\by_i = \bQ_i^H \tilde{\by} = \bR_i \bA_i \bs + \bn_i
\end{align}
where $\bQ_i^H\bQ_i=\bI_{N_L}$, $\bn_i=\sim N(\bO,\bI_{N_L})$ and $\bR_i$ is an $N_L\times N_L$ upper triangle matrix whose $(k,l)$ element is denoted by $r^{}_{kl}$. Also, $\bA_i \bs=[s_{j_1}\,s_{j_2}\,...\,s_{j_{N_L}}]^T$ with $j_{N_L} = i$ and $j_i=N_L$.
we approximate the LLR in \eqref{eqn_LLR_MaxLog} as follows:
\begin{align}
L(c_{ik})&\approx\max\limits_{s_i\in \chi_{k,1}} \eta_\text{B-Chase}(\bs) - \max\limits_{s_i\in \chi_{k,0}} \eta_\text{B-Chase}(\bs)\label{eqn_LLRmax_BChase}\\
\eta_\text{B-Chase}(\bs) &= \sum_{m=1}^{N_L} \sum_{n=1}^q b_{mn}L_a(c_{mn}) \hspace{-0.1cm}- \left|\by_i(N_L)-r^{}_{N_LN_L}s_i\right|^2 \nonumber\\ &-\sum_{l=1}^{N_L-1} \frac{\left|\by_i(l)-r^{}_{lN_L}s_i-\sum\limits_{f=l+1}^{N_L-1} r^{}_{lf} \hat{s}_\mathit{j_f^{}}-r^{}_{ll}s_{j_l}\right|^2}
{\text{var}_{\,l}} \label{eqn_LLR_BChase}
\end{align}
where $\eta_\text{B-Chase}$ is the same as $\eta_\text{max-log}$ in \eqref{eqn_LLR_MaxLog} except that the symbols $\left\{s_\mathit{j_f^{}}\right\}_{f=l+1}^{N_L-1}$ are replaced by their estimates, $\left\{\hat{s}_\mathit{j_f^{}}\right\}_{f=l+1}^{N_L-1}$, instead of being nulled out as in \eqref{eqn_I/P_O/P_linear}. Furthermore, the residual error after subtracting the symbols estimates is being accounted for using its variance $\text{var}_{\,l}$ given by
\begin{align}
\text{var}_{\,l} = 1+\sum_{f=l+1}^{N_L-1} |r^{}_{lf}|^2 \text{var}(s_\mathit{j_f^{}})
\end{align}
where the symbols estimates and variances are given by
\begin{align}
\hat{s}_\mathit{j_f^{}} &= E\left[s_\mathit{j_f^{}}\left|\left\{L_a(c_\mathit{j_f^{}n}),L_d(c_\mathit{j_f^{}n})\right\}_{n=1}^q\right.\right]\label{eqn_mean_est}\\
\text{var}(s_\mathit{j_f^{}}) &= E\left[|s_\mathit{j_f^{}}-\hat{s}_\mathit{j_f^{}}|^2\left|\left\{L_a(c_\mathit{j_f^{}n}),L_d(c_\mathit{j_f^{}n})\right\}_{n=1}^q\right.\right]\label{eqn_var_est}
\end{align}
where $|$ denotes the conditioning operator and $\left\{L_d(c_\mathit{j_f^{}n})\right\}_{n=1}^q$ are the post-detection LLRs of the bits representing the symbol $s_\mathit{j_f^{}}$ obtained as follows:
\begin{align}
L_d(c_{j_\mathit{f}n}) = \frac{\min\limits_{s_\mathit{j_f^{}}|c_\mathit{j_f^{}n}=0} \hspace{-0.05cm}\beta_\mathit{j_f^{}} \hspace{-0.1cm}-\hspace{-0.1cm} \min\limits_{s_\mathit{j_f^{}}|c_\mathit{j_f^{}n}=1} \hspace{-0.05cm}\beta_\mathit{j_f^{}}} {1+\sum_{g=f+1}^{N_L-1} |r^{}_{fg}|^2 \text{var}(s_{j_g}) },\; 2\leq f \leq N_L-1
\end{align}
where
\begin{align}
\beta_\mathit{j_f^{}} = \by_i(f) - r^{}_{fN_L}s_i-\sum_{g=f+1}^{N_L-1} r^{}_{fg} \hat{s}_{j_g}-r^{}_{ff}s_\mathit{j_f^{}}
\end{align}
We use the combined a priori and post-detection LLRs, $L_p(c_\mathit{j_f^{}n})=L_a(c_\mathit{j_f^{}n})+L_d(c_\mathit{j_f^{}n})$, and compute the mean and variance in \eqref{eqn_mean_est} and \eqref{eqn_var_est} simply as follows \cite{SIOF2010}\footnote{In most standards, the constellation mapping can be exploited to simplify the computations of the symbol mean and variance without summations}:
\begin{gather}
\hat{s}_\mathit{j_f^{}} \hspace{-0.1cm}=\hspace{-0.2cm} \sum_{s_m\in \chi} \hspace{-0.1cm}s_m \frac{\prod_{n=1}^q\left(1\hspace{-0.05cm}+\hspace{-0.05cm}(2b_{mn}-1)\text{tanh}\hspace{-0.05cm}\left(L_p(c_\mathit{j_f^{}n})/2\right)\right)}{2^q}\\
\text{var}(s_\mathit{j_f^{}})\hspace{-0.1cm}=\hspace{-0.2cm} \sum_{s_m\in \chi} \hspace{-0.1cm}|s_m|^2 \frac{\prod_{n=1}^q\left(1\hspace{-0.05cm}+\hspace{-0.05cm}(2b_{mn}-1)\text{tanh}\hspace{-0.05cm}\left(L_p(c_\mathit{j_f^{}n})/2\right)\right)}{2^q}\nonumber\\
\hspace{-5.5cm}-|\hat{s}_\mathit{j_f^{}}|^2
\end{gather}
where tanh(.) is the hyperbolic tangent function.

Next, we rewrite the maximization problem in \eqref{eqn_LLRmax_BChase} as:
\begin{gather}
\max\limits_{s_i\in \chi_{k,\hspace{-0.05cm}1(0)}} \hspace{-0.15cm} \eta_\text{B-Chase}(\bs) \hspace{-0.1cm} = \hspace{-0.2cm} \max\limits_{s_i\in \chi_{k,\hspace{-0.05cm}1(0)}} \hspace{-0.2cm} \Bigg( \underbrace{\sum_{n=1}^q b_{in}L_a(c_{in}) \hspace{-0.1cm}-\hspace{-0.1cm} \left|\bar{\by}_i(N_L)\hspace{-0.1cm}-\hspace{-0.1cm}d_is_i\right|^2}_{\triangleq \alpha_i}  \nonumber\\
+\hspace{-0.1cm}\sum_{l=1}^{N_L-1}\hspace{-0.1cm} \underbrace{\max\limits_{s_{j_l}\in \chi} \hspace{-0.1cm} \left(\sum_{n=1}^q b_{j_l n}L_a(c_{j_l n}) \hspace{-0.1cm} - \hspace{-0.1cm} \frac{\left|z_l-s_{j_l}\right|^2}
{\text{var}_{\,l}/r_{ll}^2}\right)}_{\triangleq \alpha_{i,l}}\hspace{0.05cm}\Bigg)
\label{eqn_max_problem_BChase}\\
z_l=\frac{\by_i(l)-r^{}_{lN_L}s_i-\sum_{f=l+1}^{N_L-1} r^{}_{lf} \hat{s}_\mathit{j_f^{}}}{r^{}_{ll}}\label{eqn_z_l}
\end{gather}
Then, we use our algorithm in Section \ref{sec_maximiz_comput} to solve for the maximization problem $\alpha_{i,l}$ in \eqref{eqn_max_problem_BChase}, with $E[\left|\bar{\bn}_i(l)\right|^2]$ replaced by $\text{var}_{\,l}/r_{ll}^2$ while $z_{r,l}$ and $z_{I,l}$ are replaced by the real and imaginary parts, respectively, of $z_l$ in \eqref{eqn_z_l}.

Our contribution over the B-Chase algorithm in \cite{Chase_detectorsTSP08} is two-fold. First, we account for the residual error variance in \eqref{eqn_LLR_BChase} making the detector capable of generating soft output information, unlike \cite{Chase_detectorsTSP08}, where only hard output is available. Second, we develop the algorithm to take soft input and use an efficient algorithm to solve for the maximization problems involving the a priori LLRs.


\begin{figure}
\centering
\epsfig{file=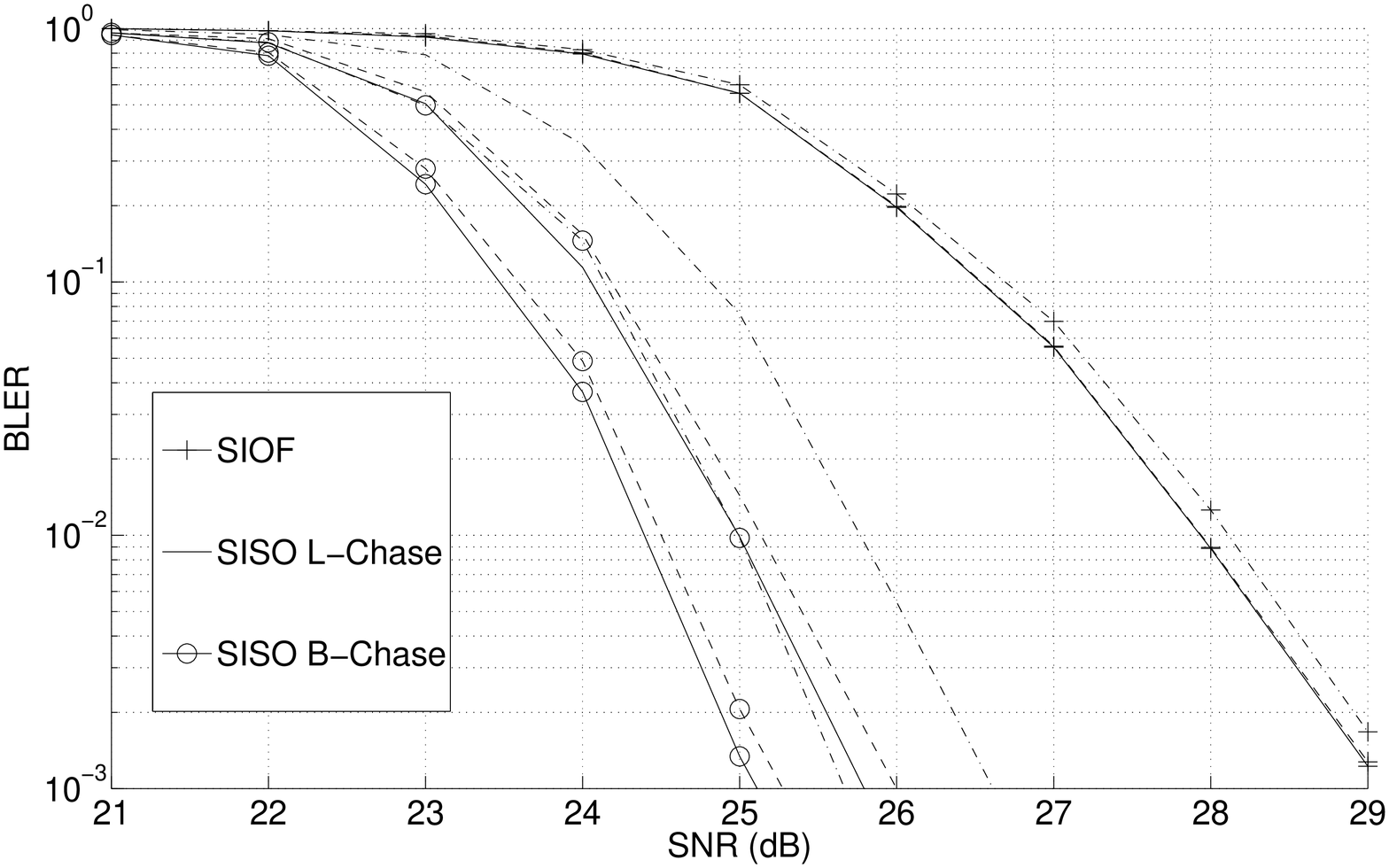,height=2.4in,width=3.45in}
\caption{Performance comparison over the first (dash-dotted lines), second (dashed lines), and third (solid lines) IDD iterations for PEDB channel with no antenna correlation and high code rate 0.83}
\label{fig_NoCorr_HighCoderate}
\end{figure}
\begin{figure}
\centering
\epsfig{file=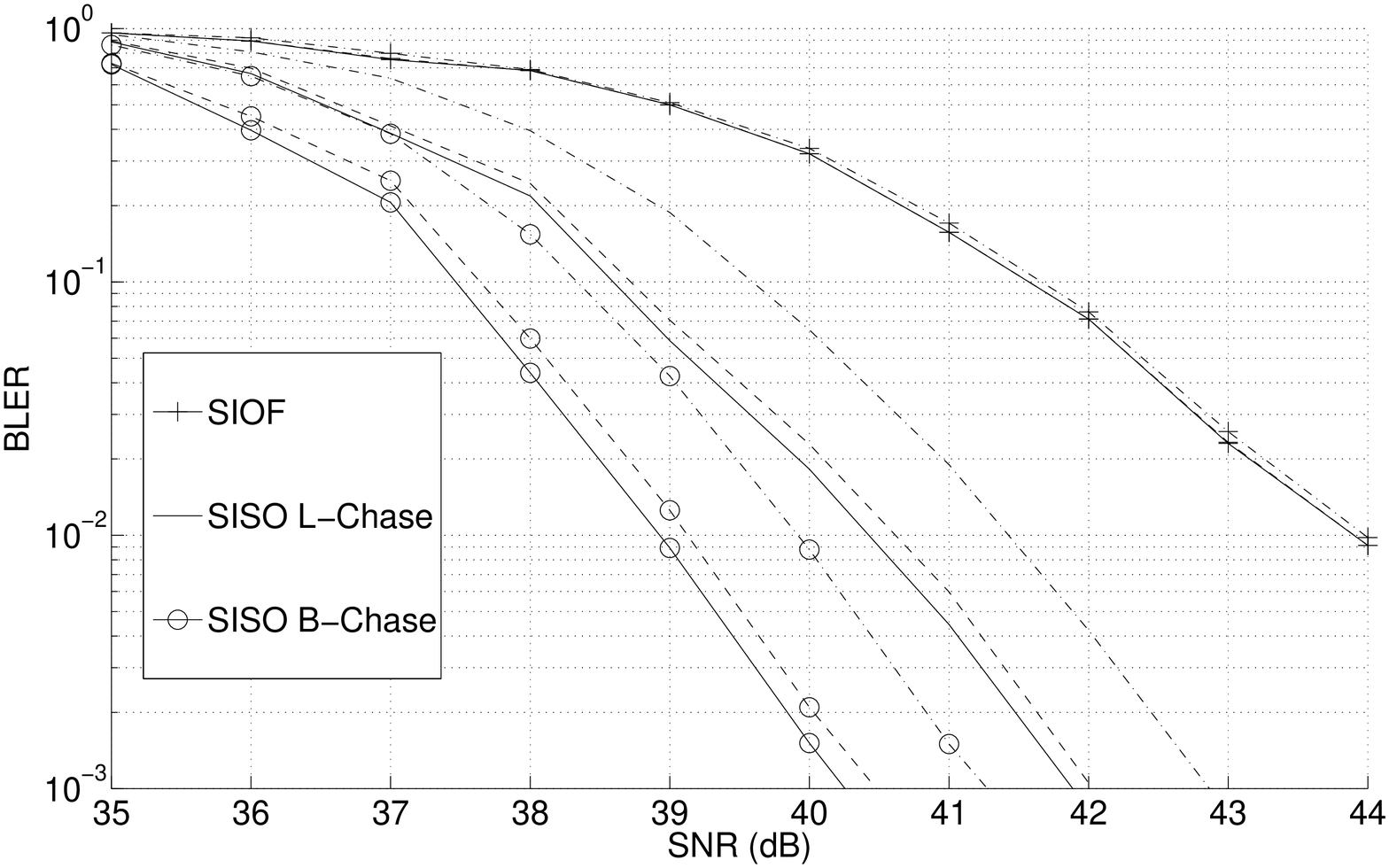,height=2.4in,width=3.45in}
\caption{Performance comparison over the first (dash-dotted lines), second (dashed lines), and third (solid lines) IDD iterations for EPA channel with high antenna correlation and low code rate 0.5}
\label{fig_HighCorr_LowCoderate}
\end{figure}

\section{Simulation Results}\label{sec_simulations}
We simulate the block error rate (BLER) performance of our proposed SISO L-Chase and SISO B-Chase algorithms, and compare them to the algorithm in \cite{SIOF2010} referred to as soft input, output, and feedback (SIOF). The performance is simulated for OFDM systems with 2048 subcarriers and 64-QAM modulation over each subcarrier for various signal-to-noise ratios (SNRs). Iterative detection and decoding with 3 iterations, is employed at the receiver, where the channel decoder is the standard LTE turbo decoder \cite{LTE_36.212}. The transmitter uses $N_t=4$ antennas to transmit $N_L=4$ layers with no precoding, i.e., $\bW=\bI_4$, and the receiver uses $N_r=4$ antennas. The performances are compared for low and high turbo code rates, namely, 0.5 and 0.83, respectively. Furthermore, two standard multi-path channel models \cite{PedB_ChannelModel} are simulated; namely, the pedestrian-B (PEDB) channel with no antenna correlation, and the extended pedestrian-A (EPA) channel with high antenna correlations, where both transmit and receive correlation coefficients are 0.9. Perfect channel knowledge is assumed at the receiver. Figs. \ref{fig_NoCorr_HighCoderate} and \ref{fig_HighCorr_LowCoderate} show that both of the proposed algorithms (SISO L-Chase and B-Chase) outperform SIOF in \cite{SIOF2010} by several dBs. The waterfall phenomenon is not observed in these figures since we simulate frequency selective channels rather than frequency flat channels. The benefit from the iterations in SIOF receiver is minimal due to the lack of coding gain when using code rate of 0.83 in the case of Fig. \ref{fig_NoCorr_HighCoderate}. High channel correlations also diminish the benefit of the SIOF method as seen in Fig. \ref{fig_HighCorr_LowCoderate} due its inability of separating out the correlated streams. Comparing Figs. \ref{fig_NoCorr_HighCoderate} and \ref{fig_HighCorr_LowCoderate}, we observe the superiority of SISO B-Chase over SISO L-Chase algorithm for channels with high antenna correlation.

\section{Conclusion}\label{sec_Conclusion}
Based on the Chase detection principle, previously known for producing hard decisions, we developed soft-input soft-output MIMO detectors for two classes of algorithms: SISO B-Chase and SISO L-Chase. An efficient method was described for using the decoder output LLRs to modulate the QAM signal constellation decision boundaries in the detector. The performances of the SISO B-Chase and SISO L-Chase were compared with the SIOF detector in \cite{SIOF2010} under different coding rates and channel conditions. Simulation results showed that the new proposed detectors have $3$ to $4$ dB advantage over SIOF at $1\%$ BLER after 3 iterations. Simulations also showed that SISO B-Chase has superior performance over SISO L-Chase for channels with high correlations.

\bibliographystyle{IEEEtran}
\bibliography{IEEEabrv,reference}

\end{document}